\documentclass[preprint,journal]{vgtc}       % preprint (journal style)
\usepackage{mathptmx}
\usepackage{graphicx}
\usepackage{times}

\usepackage{amssymb,amsfonts,amsmath}
\usepackage[noend]{algpseudocode}
\usepackage{algorithm}
\usepackage{epstopdf}

\usepackage[bookmarks,backref=true,linkcolor=black]{hyperref} %,colorlinks
\hypersetup{
  pdfauthor = {},
  pdftitle = {},
  pdfsubject = {},
  pdfkeywords = {},
  colorlinks=true,
  linkcolor= black,
  citecolor= black,
  pageanchor=true,
  urlcolor = black,
  plainpages = false,
  linktocpage
}

\onlineid{0}

\vgtccategory{Research}

\vgtcinsertpkg

\usepackage{amssymb}
\usepackage{amsmath}
\usepackage{amsthm}
\theoremstyle{plain}

\theoremstyle{definition}

\theoremstyle{remark}
\newtheorem{rem}{Remark}[section]

\title{Discovering and Visualizing Hierarchy in Multivariate Data}

\author{Kun Yang, Wing Hung Wong}
\authorfooter{
\item
 Kun Yang is a PhD student of Stanford University at Wong Lab. E-mail: kunyang@stanford.edu.
\item
 Wing Hung Wong is Stephen R. Pierce Family Goldman Sachs Professor in Science and Human Health, Professor of Statistics and Professor of Health Research and Policy at Stanford University. E-mail: whwong@stanford.edu.
}

\shortauthortitle{Yang \MakeLowercase{\textit{et al.}}: Discovering and visualizing hierarchy in the data}

\abstract{How to extract useful insights from data is always a challenge, especially if the data is multidimensional. Often, the data can be organized according to certain hierarchical structure that are stemmed either from data collection process or from the information and phenomena carried by the data itself. The current study attempts to discover and visualize these underlying hierarchies. By regarding each observation in the data as a draw from a (hypothetical) multidimensional joint density, our first goal is to approximate this unknown density with a piecewise constant function via binary partition; our non-parametric approach makes no assumptions on the form of the density. Given the piecewise constant density function and its corresponding binary partition, our second goal is to construct a connected graph and build up a tree representation of the data by level sets. To demonstrate that our method is a general data mining and visualization tool which can provide ``multi-resolution'' summaries and reveal different levels of information of the data, we apply it to two real data sets from Flow Cytometry and Social Network.} % end of abstract


\CCScatlist{ % not used in journal version
 \CCScat{K.6.1}{Management of Computing and Information Systems}%
{Project and People Management}{Life Cycle};
 \CCScat{K.7.m}{The Computing Profession}{Miscellaneous}{Ethics}
}

\begin{document}


\firstsection{Introduction}

\maketitle

%% \section{Introduction} %for journal use above \firstsection{..} instead

%With the advent of ``Big Data'' era, massive amount of data are collected and processed at an unprecedented speed. The explosion of data has been exploited in almost every field. It supports online business in commercial world, disease diagnosis and new drug development in pharmaceutical industry, governmental effort in countering terrorism and cutting-edge research in academia. The data are typically featured with high dimension and large sample size. Instead of interpreting the data through specific models, how to understand them in a human perceivable manner is a challenge. Besides the traditional tables, maps and diagrams, advanced data visualization techniques are used to reveal dynamic information in systems \cite{Bedem2013, Wilson2013}, discover connections in networks \cite{Freeman2000} and extract topology in data clouds \cite{Lum2013}.

Many data manifest certain patterns of hierarchies that are stemmed either from data collection process or from the information and phenomena carried by the data itself. Examples include census data collected at county level, state level or nation level and stem cell population differentiated into various specialized cell types. In this paper, we propose new algorithms to discover modes and visualize underlying hierarchies. We first introduce the concept of binary partitions; then develop the method to construct a class of piecewise constant density function, by regarding the multidimensional data as independent observations drawn from some hypothetical distribution. The method is motivated by the discrepancy criteria in Quasi Monte Carlo and has worst complexity $O(n\log^d n)$, where $d$ is the dimension and $n$ is the sample size. Subsequently, we introduce the tree of level sets and the algorithm to build it based on the piecewise constant density function. Through simulation and real data examples, it is shown that this binary partition based density estimate and its corresponding level-set tree provide a general tool to mine and visualize data and are capable of revealing the modes and summarizing hidden hierarchical structures.

\section{Binary Partition by Discrepancy}

\begin{figure}[ht]
\center
\includegraphics[width=.45\textwidth]{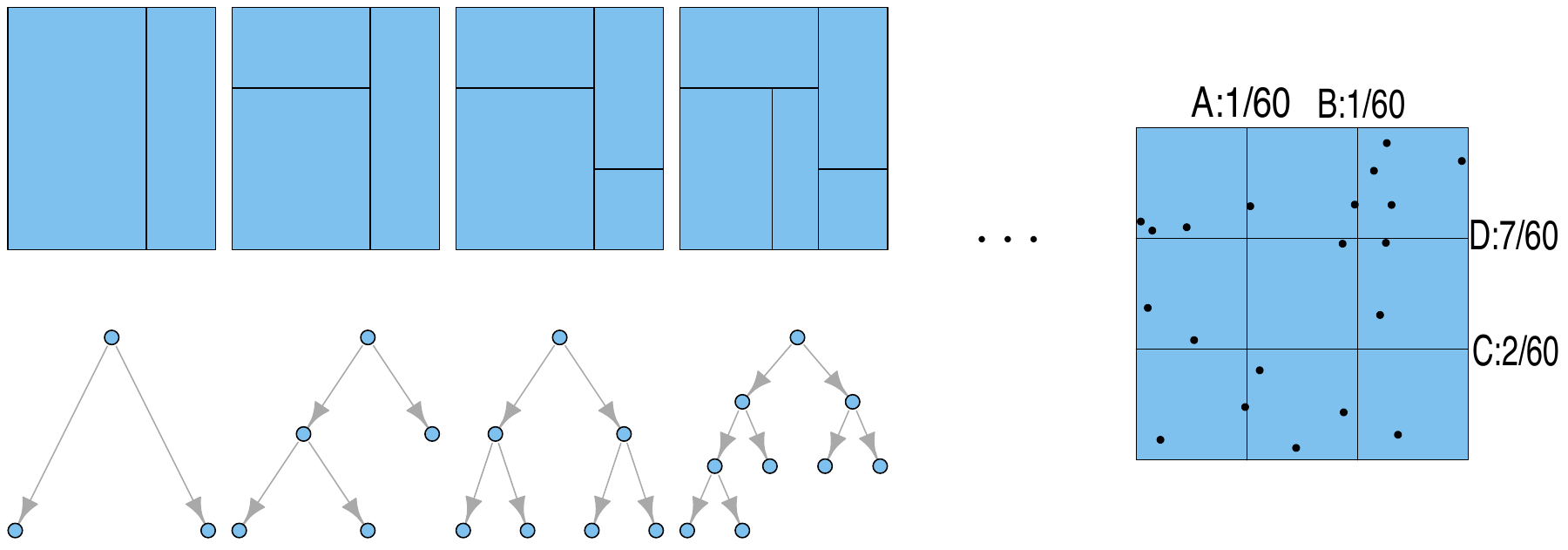}
\caption{Left: A sequence of binary partitions in two dimensional cube and the corresponding partition trees. From left to right, $t = 2, 3, 4, 5$. More information can be encoded in nodes, e.g., the dimensions and locations where the splits occur. Right:  the gaps with $m = 3$, we split the cube at location D if the hypothesis is rejected.}
\label{bp}
\end{figure}

Let $\Omega$ be a hypercube in $\mathbb{R}^d$. A binary partition $\mathcal{B}$ on $\Omega$ is a collection of sub-cubes whose union is $\Omega$. Starting with $\mathcal{B}_1 = \{\Omega\}$ at level $1$ and $\mathcal{B}_t = \{\Omega_1, \Omega_2, ..., \Omega_t\}$ at level $t$, $\mathcal{B}_{t + 1}$ is produced by dividing one of regions in $\mathcal{B}_t$ into two sub-cubes along one of its coordinates, then combining these two sub-cubes with the rest of regions in $\mathcal{B}_t$; continuing with this fashion, one can generate any binary partition in any level (Figure \ref{bp}).
\par Piecewise constant function is of fundamental importance in mathematics and statistics for its simplicity and its ability to approximate any continuous function to any degree of accuracy. In order to construct a simple yet flexible density estimator, we restrict the class of density function as the piecewise constant function on the binary partitioned sample space. Our algorithm, by exploiting the sequential build-up of binary partition, can find an optimal density estimation efficiently.
\par For piecewise constant function densities, the distribution conditioned on each piece is uniform. Thus given a binary partition, whether some of its sub-cubes needs further partitioning depends on the uniformity of the points in sub-cubes. In another word, we need to test the uniformity of points in them. Since any sub-cube is a translation and scaling of unit cube and uniformity is preserved under such transformation, it is equivalent to test the following hypothesis,
\[H_0: x\sim U[0, 1)^d, x\in \mathcal{S} = \{x_i = (x_{i1}, ..., x_{id}), x_i\in [0, 1)^d\}_{i = 1}^n \]
In the literature of quasi-Random Number Generators or quasi-Monte Carlo methods \cite{Liang2001}, there are a number of criteria for measuring whether a set of points is uniformly scattered in the unit cube $[0, 1)^d$. These criteria are called discrepancies, and they arise in the error analysis of quasi-Monte Carlo methods for evaluating integrals \cite{Owen2003}.
\par The precise definitions of the discrepancy and the variation depend on the space of integrands. For $1\leq p<\infty$, the $\mathcal{L}^p$ star discrepancy is given by
\[D_{p}^{*}(\mathcal{S}) = \Big(\int_{x\in [0, 1)^d}\Big|\frac{\#(\mathcal{S}\cap [0, x))}{n} - \prod_{j = 1}^dx_j\Big|^p\Big)^{1/p}\]
where $\#$ is the cardinality of a set. The one widely used in quasi-Monte Carlo analysis is the classic star discrepancy, i.e. $D_{1}^{*}(\mathcal{S})$. Besides $D_1^{*}$, there are $D_2^{*}$, symmetric discrepancy and centered discrepancy defined on the reproducing kernel Hilbert space, they all have interesting geometrical interpretations. One of their advantages is that their explicit formulas are available \cite{Hickernell1998}, thus, we can construct computationally tractable statistics for testing multivariate uniformity on a set of points via their formulas.
\par Discrepancy based uniformity test is shown to be more powerful than other alternatives \cite{Petrie2013}. However, if $H_0$ is rejected for a given sub-cube, a strategy of how to split the sub-cube is still required. By noting that uniformity in $[a, b] = \prod_{j = 1}^d[a_j, b_j]$ implies uniformity in each dimension, we divide $j$th dimension into $m$ equal bins $[a_j, a_j + (b_j - a_j) / m, ..., [a_j + (b_j - a_j) (m - 2) / m, a_j + (b_j - a_j) (m - 1) / m]$ for a given $m$, and keep track of the gaps at $a_j + (b_j - a_j) / m, ..., a_j + (b_j - a_j) (m - 1) / m$, where the gap $g_{jk}$ is defined as
\[g_{jk} = \Big|\frac{1}{n}\sum_{i = 1}^n\mathbf{1}(x_{ij} < a_j + (b_j - a_j) k / m) - \frac{k}{m}\Big|\] for $k = 1, ..., (m - 1)$ and $j = 1, ..., d$. Among the $(m - 1)d$ recorded gaps, we split the cube into two sub-cubes along the dimension and location corresponding to maximum gap (Figure \ref{bp}).
\par
\begin{figure}[ht]
\center
\includegraphics[width=.45\textwidth]{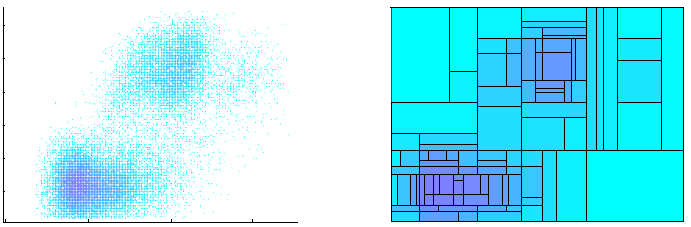}
\caption{An illustration of $p(x)$ in \eqref{eq1} in 2 dimension. Left: the point cloud; Right: the learned partition and associated densities displayed by a colormap.}
\label{ill}
\end{figure}

 As detailed in Materials and Methods section of Appenidx, the output of the density estimation is a binary partition of the sample space with associated density in each sub-cube. The density, which is a piecewise constant function, is
 \begin{equation}
 \hat{p}(x) = \sum_{i = 1}^l d(r_i)\mathbf{1}(x\in r_i)
 \label{eq1}
 \end{equation}
where $\mathbf{1}$ is indicator function; $\{r_i, d(r_i)\}_{i = 1}^l$ is the list of pairs of sub-cubes and corresponding densities (Figure \ref{ill}). Note that the number of sub-cubes is usually far less than the data size, hence $\hat{p}(x)$ provides a concise summary of the data. For a given binary tree with the partition locations encoded in each node, one can uniquely map it to a split of the sample space by recursive tree traversal. This one-to-one correspondence motivates us to utilize it as a proxy to visualize and manipulate the origin partition in high dimensions. We call this class of binary trees as ``partition trees''.
\section{Level-set Tree}
The tree of sub-level sets is widely used to represent energy or distribution landscapes \cite{Zhou2009}. A level-set tree summarizes the hierarchy among various local maxima and minima in the configuration space. Each inner node on the level-set tree is a critical level that connects two or more separate regions in the domain. Given a density function $p(x)$ on $\Omega$, define
\[\Omega_\eta^p = \{x: p(x) \geq \eta\}\]
as the level-set at level $\eta$ and $conn(\Omega_\eta^p)$ be the set of connected components. The following properties are trivial to verify,
\\
\\
\textbf{Property.} \emph{For any $0\leq\eta'<\eta$,
\begin{itemize}
  \item[1] $\forall X\in\textrm{conn}(\Omega_{\eta}^p)$, $\exists X'\in\textrm{conn}(\Omega_{\eta'}^p)$ such that $X\subseteq X'$ and $X'$ is defined as the parent of $X$.
  \item[2] $\forall X\in\textrm{conn}(\Omega_{\eta}^p)$ and $X'\in\textrm{conn}(\Omega_{\eta'}^p)$, either $X\subseteq X'$ or $X\cap X' = \emptyset$
\end{itemize}}

\begin{figure}[ht]
\center
\includegraphics[width=.3\textwidth]{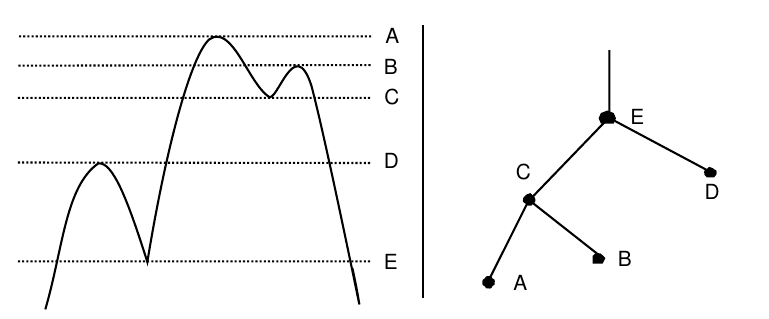}
\caption{A hypothetical density function (left) and its sub-level tree (right).}
\label{sub}
\end{figure}

For a sequence $0\leq\eta_1<\eta_2<\cdots<\eta_l$ and $\#conn(\Omega_{\eta_1}^p) = 1$, the above property de facto provides an algorithm to construct level-set tree. As illustrated in Figure \ref{sub}:  $\#conn(\Omega_\eta^p) = 1$ when $\eta < \eta_E$; $\Omega_\eta^p$ branches into two components when $\eta\in[\eta_E, \eta_D]$; $\Omega_\eta^p$ has one component again when $\eta\in(\eta_D, \eta_C)$; $\Omega_\eta^p$ splits into two smaller components at $\eta\in[\eta_C, \eta_B]$ and shrinks into one at $\eta\in(\eta_B, \eta_A]$. The corresponding level-set tree is constructed according to the parental relation defined in Property 1.

With the piecewise constant density estimation at hand, we can construct level-set tree for points instead of a given energy or density function. Unlike kernel density estimation that suffers from many local bumps and results in an overly complicated level-set tree, piecewise constant function $\hat{p}(x)$ is well suited for this purpose, partially because it smoothes out the minor fluctuations and takes only limited number of values, e.g., $l$ in \eqref{eq1}. Moreover, its simple structure makes the construction of such graph easy. According to the algorithm in \ref{sub_algo}, each sub-cube of $\hat{p}(x)$ becomes a node on level-set tree. This tree representation has merits in several aspects: i) it provides a tree visualization of the data, which is especially useful when the data are multidimensional; ii) its leaves show dense areas, i.e., modes clearly, ``mode seeking'' is a widely used technique in computer vision \cite{Comaniciu2002} and clustering; iii) it is a high level abstraction of the data and can be use to extract new features.
\\
\\
The algorithms to construct the piecewise density function and to build ``partition tree'' and ``level-set tree'' are given in Materials and Methods.

\section{Results}

We first use a simulation to illustrate the basic method and demonstrate its properties, such as the invariance to rotation and translation. Then, we apply them to two different kinds of data in two fields, namely flow cytometry and social network data, in each case discovering the relevant hierarchies.

\subsection{Simulations}
\begin{figure}[ht]
\center
\includegraphics[width=.3\textwidth]{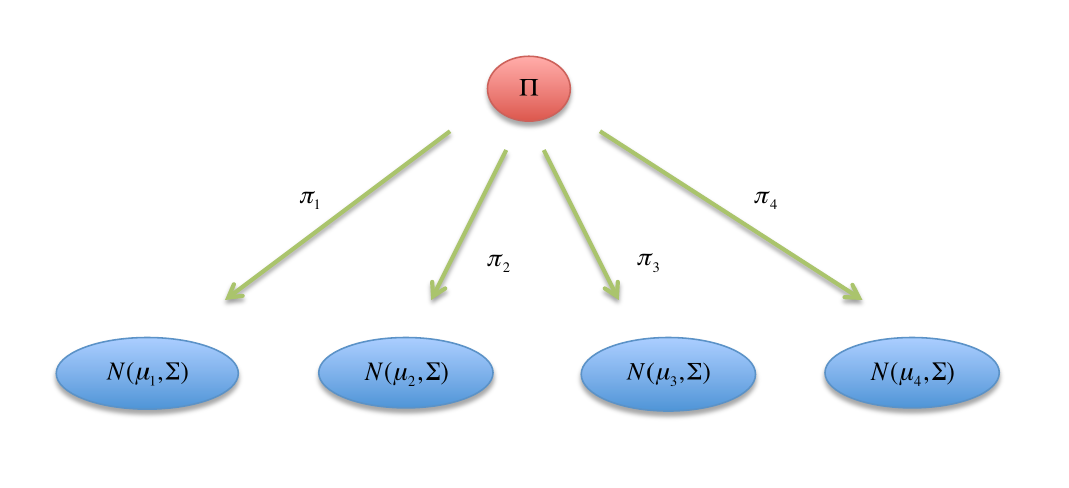}
\caption{A schematic representation of Gaussian Mixture from a generative model perspective.}
\label{flow}
\end{figure}

Consider a Gaussian mixture:
\begin{equation}\label{den}
p(x) = \sum_{i = 1}^4 \pi_i\mathcal{N}(\mu_i, \Sigma)
\end{equation}
where $(\pi_1, \pi_2, \pi_3, \pi_4) = (.25, .25, .25, .25)$ and
\begin{equation*}
  \left(\begin{array}{c}
  \mu_1\\
  \mu_2\\
  \mu_3\\
  \mu_4
  \end{array}\right) = \left(\begin{array}{rrrrrr}
  2 & 2 &  & \cdots\\
  -2 & 2 &  & \cdots\\
   & -2 & 2 & \cdots\\
   & -2 & -2 & \cdots
  \end{array}\right)_{4\times 10}
\end{equation*}
and
\begin{equation*}
  \Sigma = \left(\begin{array}{rrrrrr}
  1 & .1 &  &  &  & \cdots\\
  .1 & 1 & .1 &  & & \cdots\\
   & .1 & 1 & .1 & & \cdots\\
   &  & .1 & 1 & .1 & \cdots\\
   &  &  & .1 & 1 & \cdots\\
  \vdots & \vdots & \vdots & \vdots & \ddots & \vdots\\
   &  &  & .1 & 1 & .1\\
   & & & & .1 & 1

  \end{array}\right)_{10\times 10}
\end{equation*}
where void entries are 0s. From a generative model perspective \cite{Bishop2006}, the data generation process can be represented schematically as in Figure \ref{flow}: the cluster index is sampled according to $\mathbf{\pi}$, then $x$ is sampled from corresponding Gaussian distribution.

\begin{figure}[ht]
\centerline{\includegraphics[width=0.5\textwidth]{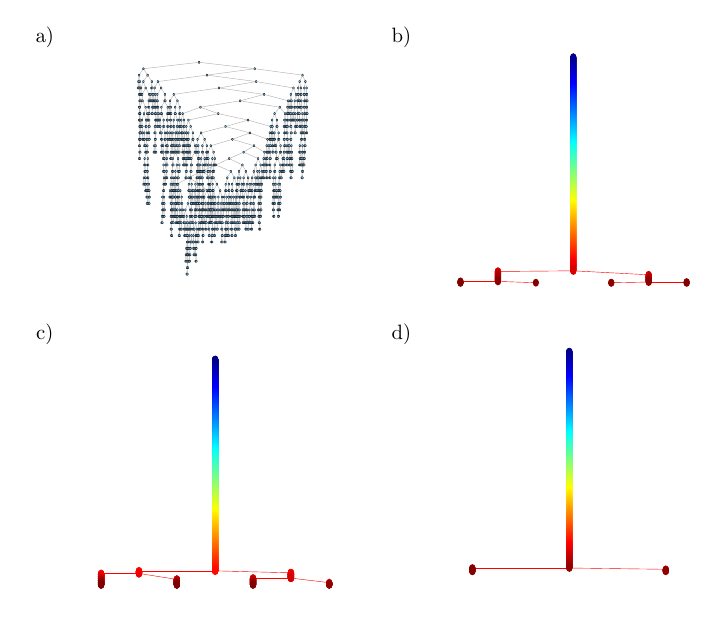}}
\caption{Partition tree and sub-level trees for samples generated from the Guassian Mixture, the colors from blue to red on SLT represents the average densities from low to high as defined in \eqref{aveden}: a) partition tree; b) corresponding sub-level tree; c) sub-level tree of the rotated and translated samples; d) sub-level tree after trimming down the highest 5 levels.}
\label{simsub}
\end{figure}

50,000 samples are drawn from \eqref{den}, we use our methods to let the data ``speak'' for itself, i.e., to recover the hierarchy in Figure \ref{flow}. The partition tree and level-set tree are shown in Figure \ref{simsub}.ab. It is clear that the four branches of the level-set tree in Figure \ref{simsub}.b correspond to the four clusters in $p(x)$. Moreover, richer information is available from the trees: the two sub-branches indicate the fact that cluster 1, 2 and cluster 3, 4 are closer to each other, because they merge before the four clusters becoming one. In fact, as we trim down the highest 5 levels of partition tree, only the sub-branches are visible, as shown in Figure \ref{simsub}.d. Figure \ref{simsub}.c demonstrates the invariant of level-set tree under rotation and translation. In a word, without knowing the distribution a priori, the hierarchy in the data is revealed by our methods.
\subsection{Real Data}
\subsubsection{Flow Cytometry}
Multi-parameter flow cytometry allows to measure multiple characteristics of single cells simultaneously; it provides insights into cellular differentiation, cellular hierarchy and disease diagnostics. Despite the increase in throughput and the number of parameters per single cell, there are limited number of methods for visualizing and analyzing multidimensional single-cell data. Moreover, cell differentiation creates the underlying hierarchy among the cell populations. Traditional clustering algorithms are capable of finding mature cell populations (heterogeneity), whereas they ignore the continuity of phenotypes. As an attempt to capture this important aspect in cell populations, we apply our methods to the mouse bone marrow data studied in \cite{Qiu2011}.

We regard each cell as one sample in the sample space, i.e., if there are $d$ markers attached to a single cell, then the whole data set is generated from a hypothetical $d$ dimensional distribution. Mature cell populations concentrate in some high density areas, i.e., the modes or local maxima on the domain. By learning the $d$ dimensional density and constructing the affiliated level-set trees, each cell population is clustered around the set of sub-cubes in each branch of the level-set tree. Based on the expression levels of markers in these populations, we can infer their hierarchy accordingly.

One practical issue needs to be addressed for most of the Cytometry analysis techniques: there is asymmetry in sub-populations; by optimizing a predefined loss function, it is possible that some sparse yet crucial populations are overlooked if the algorithms take most of the efforts to control the loss in denser areas. A remedy for this issue is to perform a down-sampling \cite{Aghaeepour2013, Qiu2011} step to roughly equalize the densities among populations then up-sampling after populations are identified. However, this step is dangerous that it may fails to sample enough cells in sparse populations, as a result, these populations are lost in the down-sampled data. In contrast, our approach does not require down-sampling step, and the asymmetry among populations are captured by the densities in sub-cubes.

\begin{figure*}[ht]
\center
\includegraphics[width=.6\textwidth]{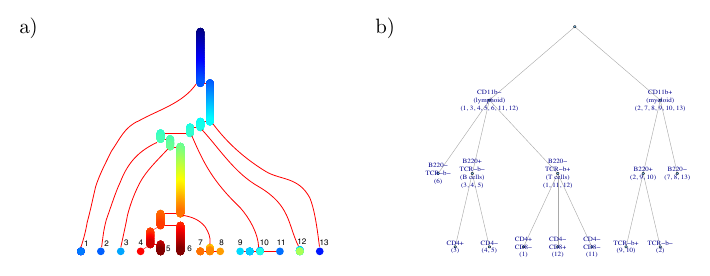}
\caption{Mouse bone marrow: a) sub-level tree learned from 8 markers, CD11b, B220, TCR-$\beta$, CD4, CD8, c-kit, Sca-1; b) corresponding cellular hierarchy built from the expression levels of markers in each sub-populations according to the marker sequence: CD11b, B220, TCR-$\beta$, CD4, CD8.}
\label{mouse}
\end{figure*}

For the mouse bone marrow data, we choose the 8 markers (SSA-C, CD11b, B220, TCR-$\beta$, CD4, CD8, c-kit, Sca-1) that are relevant to the cell types of interests; the number of cells is $\thicksim$380,000 after removing mutli-cell aggregates and co-incident events. As shown in Figure \ref{mouse}, 13 sub-populations are identified (\cite{Qiu2011} and its supplementary materials). We can arrange them into a hierarchical dendrogram: at first level, they are grouped by expression levels of CD11b; subsequently, the CD11b- sub-populations are grouped according to B220 and TCR-b then further splitted according to CD4 and CD8 on the next level; the CD11b+ sub-populations are grouped by B220 then by TCR-b.
%given the sequence CD11b $\rightarrow$ B220 $\rightarrow$ TCR-$\beta$ $\rightarrow$ CD4 $\rightarrow$ CD8, the populations are grouped according to their positive or negative expression of each marker; the hierarchy in Figure \ref{mouse}.b can be constructed accordingly.

\subsubsection{Community structure in Social Networks}

Diverse systems in various fields take the form of networks. In this study, we consider the community property which is found in many real networks such as social networks, bio networks and technological networks. In this example, we offer another approach to visualize the structure of the network by our sub-level tree algorithm. Analogues to the hierarchical clustering \cite{Zhang2007}, it is a tree representation; however, it is much sparser and reveals the communities in its branches. We demonstrate that our methods can be used to detect the communities and reveal their denseness (cohesiveness) and discover the ``transitional'' nodes between the communities.

Given an undirected, unweighted $n-$vertices graph (network) $G =(V,E)$. The Laplacian matrix is defined as
\begin{equation}
  L_{ij} =  \left\{
  \begin{array}{rl}
  1, & i\sim j\\
  -d_i, & i = j\\
  0, & i \nsim j
  \end{array}\right.
\end{equation}
where $i\sim j$ ($i\nsim j$) means that the $i$th and $j$th vertices are (not) adjacent, and $d_i$ is the degree of the vertex. In the spectral methods of graph clustering \cite{Qin2013}, we select the leading $d$ eigenvectors of $L$ (or regularized $L$ \cite{Chaudhuri2012}) and apply the $k-$means clustering algorithm on the $n$ $d-$dim vectors. Since clusters found by $k-$means are related to the modes of the underlying distribution, level-set tree can be used to ``display'' these modes. Thus, we replace $k-$means by level-set tree algorithm instead. The vertices represented by these $d-$dim vectors are contained in sub-cubes. All the vertices belonging to the set of sub-cubes of a level-set tree's branch correspond to a community. However, some vertices are not contained in the sub-cubes on the branches, we define them as a ``transitional'' vertices since it plays a key role in the formation of communities.

\begin{figure}[ht]
\centerline{\includegraphics[width=.5\textwidth]{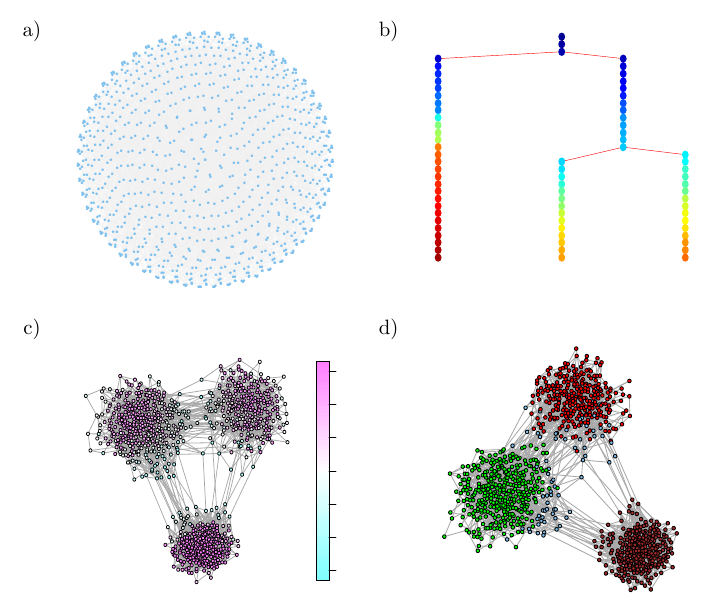}}
\caption{Network: a) original network plotted on the sphere; b) corresponding sub-level tree of the 3 leading eigenvectors of the Laplacian; the two closer communities merge first, then merge with the third one; c) vertices in the network are colored according to densities; d) communities colored by red, brown and green, the transitional vertices are colored by blue.}
\label{network}
\end{figure}

We simulate a 1,000 vertices network and define the adjacency matrix $M$ as follows: 1) Assign $M_{i, i + 1} = 1, i = 1, 2, ..., 999$ to make the network connected; 2) Construct three communities: $A = \{1, ..., 300\}$, $B = \{301, ..., 600\}$, $C = \{601, ..., 1,000\}$ with the edges in each community assigned as: i) $M_{i, j} = 1, i, j \in A$ with probability 0.01; ii) $M_{i, j} = 1, i, j \in B$ with probability 0.02; iii) $M_{i, j} = 1, i, j \in C$ with probability 0.008; and the edges between communities assigned as: i) $M_{i, j} = 1, i\in A, j\in B$ with probability 0.0001; ii) $M_{i, j} = 1, i\in B, j\in C$ with probability 0.0001; iii) $M_{i, j} = 1, i\in C, j\in A$ with probability 0.0005. We use the 3 leading eigenvectors of $L$ to learn a binary partition and sub-level tree. In Figure \ref{network}, the three communities are identified on the branches of level-set tree; since $A$ and $C$ are closer to each other, their corresponding branches on level-set tree merge first.

\begin{figure*}[ht]
\centerline{\includegraphics[width=1.\textwidth]{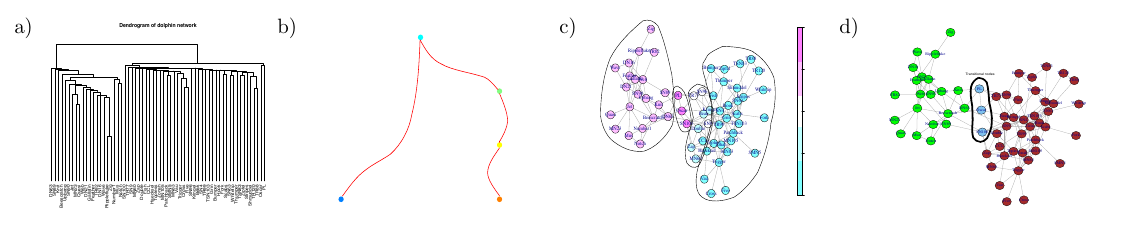}}
\caption{Dolphin Social Network: a) the dendrogram by hierarchical clustering according to \cite{Zhang2007}; b) sub-level tree of the leading 2 eigenvectors of the Laplacian; c) the dolphin network that vertices are colored and grouped according to densities; d) two dolphin communities and the ``transitional'' vertices, particularly, SN100, which plays an important role in the formation of communities, is identified.}
\label{dolphin}
\end{figure*}

We also apply our methods to classic dolphin social network \cite{Lusseau2003, Lusseau2004}, it was constructed from observations of a community of 62 bottlenose dolphins over a period of seven years between 1994 and 2001. The two communities are correctly identified as shown in Figure \ref{dolphin}.c and the relative ``densities''(cohesiveness) of both communities are also colored in Figure \ref{dolphin}.b. More interestingly, SN100, the individual with the highest connectivity in both communities and playing an important role in the fission and reunion of the dolphin community, are identified as a ``transitional'' vertex.


\section{Discussion}

Complex data can be understood in different perspectives. Classic methods display simple statistics such as mean, variance or point clouds with dimension no more than 3; early attempts to visualize high dimensional data, such as Chernoff faces \cite{Chernoff1973}, are applicable to relatively small data sets. As data size increases, focus has been shifted to the sparse representations, e.g., \cite{Lum2013} tries to capture data topology (``shape'') and summarize it in a graph.

Our methods are designed to mine another aspect of the data---modes and hierarchies. They are non-parametric and unsupervised in nature, thus they do not suffer from the bias of specific model or assumptions. In Results, we show that they are applicable to different types of problem. Another possible direction is to build a mode seeking algorithm based on our level-set tree and apply it to image segmentation.

\appendix
\section{Materials and Methods}
\subsection{Binary Partition by Discrepancy}
$P(\cdot)$ defines the set of points and $\Pr(\cdot)$ defines the probability mass in a sub-cube respectively. Without loss of generality, we assume that $\Omega = [0, 1)^d$ and $P(\Omega) = \{x_i = (x_{i1}, x_{i2}, ..., x_{id})\}_{i = 1}^n$.
\begin{algorithmic}[1]
\Procedure{density-estimator}{$\Omega, m, \alpha$}
\State $T = \{\Omega\}$, $\Pr(\Omega) = 1$
\While{true}
    \State $\tilde{T} = \emptyset$
    \For{each $r = \prod_{j = 1}^d[a_j, b_j]\in T$}
        \State Transform $P(r) = \{x_{r_j}\}_{j = 1}^{n_r}$ to
        \State $\quad\quad\tilde{P}(r) = \{\tilde{x}_{r_j} = (\frac{x_{r_j, 1} - a_{1}}{b_{1} - a_{1}}, ..., \frac{x_{r_j, d} - a_{d}}{b_{d} - a_{d}})\}_{j = 1}^{n_r}$
        \State Test the uniformity of $\tilde{P}(r)$ by discrepancy
        \State Split $r$ into $\{r_1, r_2\}$ along the maximum gap $\{g_{jk}\}$
        \If{$r$ is divided}
        \State $\tilde{T} = \tilde{T}\cup\{r\}$
        \State Continue
        \EndIf
        \For{$j\gets 1, ..., d$}\Comment count the points in each bin
            \State $B_i = 0, i = 1, 2, ..., m$
            \For{each $x = (x_1, ..., x_d)\in\tilde{P}(r)$}
                \State $B_{\lfloor mx_j\rfloor + 1} = B_{\lfloor mx_j\rfloor + 1} + 1$\\
            \EndFor
            \State Record the gaps as $\{g_{jk}\}_{k = 1}^{m - 1}$
        \EndFor
        \If{$r$ is divided}
        \State $\tilde{T} = \tilde{T}\cup\{r_1, r_2\}$
        \State $\textrm{Pr}(r_1) = \textrm{Pr}(r)\frac{\#P(r_1) + \alpha}{\#P(r) + 2\alpha}$
        \State $\textrm{Pr}(r_2) = \textrm{Pr}(r) - \textrm{Pr}(r_1)$
        \EndIf
    \EndFor
    \If{$\tilde{T} == T$}
        \State \textbf{return} T
    \Else
        \State $T = \tilde{T}$
    \EndIf
\EndWhile
\EndProcedure
\end{algorithmic}
\begin{rem}
  The partition tree can be constructed as a byproduct by bookkeeping the parental relations of partitions.
\end{rem}
The density in $r$ is recovered by $d(r) = \textrm{Pr}(r) / |r|$, where $|r|$ is the volume of $r$; $\alpha > 0$ is a Laplace smoother (pseudo count). In line 8, we test the uniformity hypothesis by the symmetric discrepancy \cite{Liang2001} as follows, let
\[A = \frac{1}{n}\sum_{i = 1}^n\prod_{j = 1}^d(1 + 2x_{ij} - 2x_{ij}^2)\]
\[B = \frac{2^{d - 1}}{n(n - 1)}\sum_{i < j}\prod_{k = 1}^d(1 - |x_{ik} - x_{jk}|)\]
\[C = (4/3)^d, \eta = (9/5)^d - (6/9)^d\]
then
\begin{equation}
\sqrt{n}[(A - C) + 2(B - C)]/(5\sqrt{\eta}) \xrightarrow{\mathcal{D}}\mathcal{N}(0, 1)
\label{htest}
\end{equation}
Note that $B$ can be computed in $O(n\log^{d - 1}n)$ according to Frank and Heinrich's algorithm \cite{Doerr2013}; at level $t$, the number of samples in each sub-cube is $n_i, i = 1, ..., t$, the complexity is \[\sum_{i = 1}^t n_i\log^{d - 1}n_i\leq \sum_{i = 1}^tn_i\log^{d - 1}n = n\log^{d - 1}n\]
thus, the total complexity is $O(l\cdot n\log^{d - 1}n)$, where $l$ is the deepest level, which is a moderate number in our experience and can be specified by the user as well.
\par As shown in \cite{Liang2001}, discrepancy based test is powerful even when the sample size is less than 1,000. We can also compute \eqref{htest} by sub-sampling (say 500 points, which works very well in our examples). Since there are $t$ sub-cubes in level $t$ and the uniformity test in each sub-cube takes $O(m\log^{d - 1}m)$ with $m$ samples,  the complexity is at most
\[\sum_{t = 1}^l O(m\log^{d - 1}m)t = O(m\log^{d - 1}m\cdot l^2)\]
\par Both computing strategies yield similar empirical results, but the $O(m\log^{d - 1}m\cdot l^2)$ one becomes attractive when the data size is large.

\subsection{Graph of the Partition}
For a given binary partition $\mathcal{B}$ and the list of pairs of sub-cubes and corresponding densities $\{r_i, d(r_i)\}_{i = 1}^l$ as in \eqref{eq1}, we build a graph $G$ based on the adjacency of sub-regions and each sub-cube is a node on the graph. The algorithm to determine the adjacency of sub-region $i, j$ is:
\begin{algorithmic}[1]
\Procedure{is-adjacent}{$r_i,r_j$}
\State $c_k = (c_{k1}, ..., c_{kd})$: the center of $r_k, k\in \{i, j\}$
\State $l_k = (l_{k1}, ..., l_{kd})$: the width of $r_k$ in each dimension, $k\in\{i, j\}$
\For{$k\gets 1, ..., d$}
    \If{$|c_{ik} - c_{jk}| > (l_{ik} + l_{jk}) / 2$}
        \State \textbf{return} False
    \EndIf
\EndFor
\State \textbf{return} True
\EndProcedure
\end{algorithmic}
$G$ is constructed by connecting adjacent sub-cubes.

\subsection{Level-set Tree}\label{sub_algo}
The complete description of the algorithm is:
\begin{algorithmic}[1]
\State Input: $\mathcal{B}, \Pr(\cdot)$
\State Output: Level-Set Tree $\mathcal{T}$
\Procedure{level-set-tree}{$\mathcal{B}, \Pr(\cdot)$}
\State $t$: the number of sub-cubes (i.e., levels) in $\mathcal{B}$\;
\State $r_{(1)}, ..., r_{(t)}$: the sub-cubes in $\mathcal{B}$ ordered decreasingly by $d(r_i), i = 1, ..., t$\;
\State $G$: the graph of sub-cubes by \textsc{is-adjacent}($r_i,r_j$);
\State $G[r_{(1)}, ..., r_{(i)}]$: the sub-graph induced by $[r_{(1)}, ..., r_{(i)}]$ and $G(\emptyset) = \emptyset$\;
\State $\Xi_0, \Xi_1, \cdots$ : $\Xi_i = [C_{i_1}, ..., C_{i_a}]$ is the set of connected components of sub-graph induced by $G[r_{(1)}, ..., r_{(i)}]$\;
\State $\pi(\cdot)$: the most recent added sub-rectangle in a connected component\;
\State $\Pi_0, \Pi_1, \cdots$ : $\Pi_i = [\pi(C_{i - 1}), ..., \pi(C_{i_a})]$, where $\Xi_i = [C_{i_1}, ..., C_{i_a}]$\;
\State $\wp(\cdot)$ : the parent of each sub-cube in $\mathcal{B}$\;
\State $\textrm{Color}(\cdot)$ : the color of each-cube in $\mathcal{B}$\;
\State $\Xi_0 = \emptyset$\;
\State $\pi_0 = \emptyset$\;
\For{$k\gets 1$ to $t$}
    \If{$r_{(k)}$ is adjacent to $[C_{1}, C_{2}, ..., C_{m}]_{m\geq 1}\subseteq\Xi_{k - 1}$}
    \State $\Xi_{k} = \{r_{(k)}\cup[C_{1}, C_{2}, ..., C_{m}]_{m\geq 1}, \Xi_{k - 1}\backslash [C_{1}, C_{2}, ..., C_{m}]_{m\geq 1}$\}\;
    \State $\Pi_k = \{r_{(k)}, \Pi_{k - 1}\backslash\cup [\pi(C_1), \pi(C_2), ..., \pi(C_m)]$\}\;
    \State $\wp(\pi(C_i)) = r_{(k)}, i = 1, ..., m$\;
    \State Color$(r_{(k)})$ = average density($r_{(k)}\cup[C_{1}, C_{2}, ..., C_{m}]_{m\geq 1}$)\;
    \Else
    $\Xi_{k} = [\Xi_{k - 1}, r_{(k)}]$\;
    $\Pi_{k} = [\Pi_{k - 1}, r_{(k)}]$\;
    Color$(r_{(k)})$ = average density$(r_{(k)})$\;
    \EndIf
\EndFor
\State $\mathcal{T}$ is build via $\wp$, Color($\cdot$)\;
\State \textbf{return} $\mathcal{T}$\;
\EndProcedure
\end{algorithmic}
Starting with empty set $\Xi_0$ at step 0, the sub-rectangle is added into $\Xi$ sequentially according to the decreasing order of densities. At $k$th step, we have the induced sub-graph $G[r_{(1)}, ..., r_{(k - 1)}]$ and its connected components $\Xi_{k - 1}$. There are two scenarios when $r_{(k)}$ is added into $\Xi_{k - 1}$: i) $r_{(k)}$ is adjacent to multiple components $[C_{1}, C_{2}, ..., C_{m}]_{m\geq 1}$, then $\Xi_{k} = \{r_{(k)}\cup[C_{1}, C_{2}, ..., C_{m}]_{m\geq 1}, \Xi_{k - 1}\backslash [C_{1}, C_{2}, ..., C_{m}]_{m\geq 1}$\} and $r_{(k)}$ is the parent of $[\pi(C_{1}), \pi(C_{2}), ..., \pi(C_{m})]_{m\geq 1}$; ii) $r_{(k)}$ is disconnected with all the components in $\Xi_{k - 1}$, then $\Pi_{k} = \{\Pi_{k - 1}, r_{(k)}\}$ and $r_{(k)}$ is a leaf.

 At each step, we also keep track of the average density in each component; the average density is defined as the ratio between the total mass and total volume in the component, i.e., the average density of $g$ is
\begin{equation}
\textrm{average density}(g) = \frac{\sum_{r\in g}|r|d(r)}{\sum_{r\in g}|r|}
\label{aveden}
\end{equation}
The tree nodes can be colored according to the average density when the sub-region is included in $S$ for the first time.
\acknowledgments{Kun Yang is supported by General Wang Yaowu Stanford Graduate Fellowship and The Simons Math+X fellowship; Wing Hung Wong is supported by NSF grants DMS 0906044 and 1330132.}

\bibliographystyle{abbrv}
%%use following if all content of bibtex file should be shown
%\nocite{*}
\bibliography{PNASTMPL}

\begin{thebibliography}{10}

\bibitem{Aghaeepour2013}
N.~Aghaeepour, G.~Finak, H.~Hoos, T.~R. Mosmann, R.~Brinkman, R.~Gottardo,
  R.~H. Scheuermann, F.~Consortium, and D.~Consortium.
\newblock Critical assessment of automated flow cytometry data analysis
  techniques.
\newblock {\em Nature methods}, 2013.

\bibitem{Bishop2006}
C.~M. Bishop and N.~M. Nasrabadi.
\newblock {\em Pattern recognition and machine learning}, volume~1.
\newblock springer New York, 2006.

\bibitem{Chaudhuri2012}
K.~Chaudhuri, F.~C. Graham, and A.~Tsiatas.
\newblock Spectral clustering of graphs with general degrees in the extended
  planted partition model.
\newblock {\em Journal of Machine Learning Research-Proceedings Track},
  23:35.1--35.23, 2012.

\bibitem{Chernoff1973}
H.~Chernoff.
\newblock The use of faces to represent points in k-dimensional space
  graphically.
\newblock {\em Journal of the American Statistical Association},
  68(342):361--368, 1973.

\bibitem{Comaniciu2002}
D.~Comaniciu and P.~Meer.
\newblock Mean shift: A robust approach toward feature space analysis.
\newblock {\em Pattern Analysis and Machine Intelligence, IEEE Transactions
  on}, 24(5):603--619, 2002.

\bibitem{Doerr2013}
C.~Doerr, M.~Gnewuch, and M.~Wahlstr\'{o}m.
\newblock Calculation of discrepancy measures and applications.
\newblock {\em Preprint}, 2013.

\bibitem{Hickernell1998}
F.~Hickernell.
\newblock A generalized discrepancy and quadrature error bound.
\newblock {\em Mathematics of Computation of the American Mathematical
  Society}, 67(221):299--322, 1998.

\bibitem{Liang2001}
J.-J. Liang, K.-T. Fang, F.~Hickernell, and R.~Li.
\newblock Testing multivariate uniformity and its applications.
\newblock {\em Mathematics of Computation}, 70(233):337--355, 2001.

\bibitem{Lum2013}
P.~Lum, G.~Singh, A.~Lehman, T.~Ishkanov, M.~Vejdemo-Johansson, M.~Alagappan,
  J.~Carlsson, and G.~Carlsson.
\newblock Extracting insights from the shape of complex data using topology.
\newblock {\em Scientific reports}, 3, 2013.

\bibitem{Lusseau2004}
D.~Lusseau and M.~E. Newman.
\newblock Identifying the role that animals play in their social networks.
\newblock {\em Proceedings of the Royal Society of London. Series B: Biological
  Sciences}, 271(Suppl 6):S477--S481, 2004.

\bibitem{Lusseau2003}
D.~Lusseau, K.~Schneider, O.~J. Boisseau, P.~Haase, E.~Slooten, and S.~M.
  Dawson.
\newblock The bottlenose dolphin community of doubtful sound features a large
  proportion of long-lasting associations.
\newblock {\em Behavioral Ecology and Sociobiology}, 54(4):396--405, 2003.

\bibitem{Owen2003}
A.~B. Owen.
\newblock Quasi-monte carlo sampling.
\newblock {\em Monte Carlo Ray Tracing: Siggraph}, pages 69--88, 2003.

\bibitem{Petrie2013}
A.~Petrie and T.~R. Willemain.
\newblock An empirical study of tests for uniformity in multidimensional data.
\newblock {\em Computational Statistics \& Data Analysis}, 2013.

\bibitem{Qin2013}
T.~Qin and K.~Rohe.
\newblock Regularized spectral clustering under the degree-corrected stochastic
  blockmodel.
\newblock In {\em Advances in Neural Information Processing Systems}, pages
  3120--3128, 2013.

\bibitem{Qiu2011}
P.~Qiu, E.~F. Simonds, S.~C. Bendall, K.~D. Gibbs~Jr, R.~V. Bruggner, M.~D.
  Linderman, K.~Sachs, G.~P. Nolan, and S.~K. Plevritis.
\newblock Extracting a cellular hierarchy from high-dimensional cytometry data
  with spade.
\newblock {\em Nature biotechnology}, 29(10):886--891, 2011.

\bibitem{Zhang2007}
S.~Zhang, X.-M. Ning, and X.-S. Zhang.
\newblock Graph kernels, hierarchical clustering, and network community
  structure: experiments and comparative analysis.
\newblock {\em The European Physical Journal B}, 57(1):67--74, 2007.

\bibitem{Zhou2009}
Q.~Zhou and W.~H. Wong.
\newblock Energy landscape of a spin-glass model: Exploration and
  characterization.
\newblock {\em Physical Review E}, 79(5):051117, 2009.

\end{thebibliography}
\end{document}